\documentclass[aps,prb,reprint,groupedaddress]{revtex4-1}

\usepackage{amsmath,amssymb} % mathematics
\usepackage{amsmath} % mathematics
\usepackage{amsfonts} % mathematics
\usepackage{bm} % bold in math-mode
\usepackage[pdftex]{graphicx} % figure
\usepackage{braket} % braket

\begin{document}

\title{Entanglement spectrum as a marker for phase transitions in the density embedding theory for 
interacting spinless fermionic models}
\author{Xavier Plat}
\affiliation{Condensed Matter Theory Laboratory, RIKEN, Wako, Saitama 351-0198, Japan}
\author{Chisa Hotta}
\affiliation{Graduate School of Arts and Sciences, The University of Tokyo, 3-8-1, Komaba, Meguro-ku, Tokyo 1538902, Japan. }
 %\footnote{\vspace*{-10mm} electric address: chisa@phys.c.u-tokyo.ac.jp} 

\date{\today}
%%{submitted in }
\begin{abstract}
Entanglement related properties work as nice fingerprint of the quantum many-body wave function. 
However, those of fermionic models are hard to evaluate in standard numerical methods because they suffer from finite size effects. 
We show that a so-called density embedding theory (DET) 
can evaluate them without size scaling analysis in comparably high quality with those
obtained by the large-size density matrix renormalization group analysis. 
This method projects the large scale original many-body Hamiltonian to the small number of basis sets defined on a local cluster, 
and optimizes the choice of these bases by tuning the local density matrix. 
The DET entanglement spectrum of one-dimensional interacting fermions perfectly reproduces the exact ones and works as a marker of the phase transition point. 
It is further shown that the phase transitions in two-dimension could be determined by 
the entanglement entropy and the fidelity that reflects the
change of the structure of the wave function.
\end{abstract}
\maketitle

%*%*%*%*%*%*%*%*%*%*%*%*%*%*%*%*%*%*%*%*%*%*%*%*%*%*
%*%*%*%*%*%*%*%*%*%*%*%*%*%*%*%*%*%*%*%*%*%*%*%*%*%*
Historical findings in condensed matter physics owe much to the identification of the wave functions 
of the corresponding new phases of matter, as exemplified by the Bardeen-Cooper-Schrieffer wave function of the superconductivity\cite{BCS1957} 
and the Laughlin wave function for the fractional quantum Hall effect\cite{Laughlin1983}. 
In more recent studies, it turned out that direct access to the ground state wave function has the advantage 
in that a fidelity and entanglement-related properties can be straightforwardly evaluated. 
These quantities, originally coming from the quantum information field, have proven to be useful for the study of strongly correlated systems. The fidelity is defined as the magnitude of the overlap between two wave functions. Considering two ground states corresponding to slightly different Hamiltonian parameters, a quantum phase transition is detected by a drop in the fidelity\cite{Zanardi2006}, which can also contain details about the nature of the transition~\cite{Cozzini2007,Albuquerque2010,Gu2010}. 
The advantage of the fidelity approach is that it does not require knowledge of an order parameter, while still providing a clear sign of a transition. Similarly, entanglement is shown to contain information about the phase diagram in general~\cite{Amico2008,Eisert2010}. In particular, several types of entanglement measurements are expected to be valuable probes for quantum phase transitions
~\cite{Legeza2006,Chepiga2016,Hatsugai2006}.

In the present paper, we use density embedding theory~\cite{Bulik2014a} (DET), a variation of the density matrix embedding theory method (DMET) originally introduced in Ref.~[\onlinecite{Knizia2012}], as a systematic and efficient method to simulate fermionic systems with inter-site two-body interactions. The method relies on a wave function based embedding procedure, in which an impurity model of a much smaller size than the original bulk problem is constructed and solved. The method eventually yields a wave function which is expected to reproduce bulk properties.

Let us first outline the main ideas of DET. We first divide a lattice of $N$ sites into a small subsystem called impurity fragment of size $N_{\mathrm{imp}}$ and its complement of size $N-N_{\mathrm{imp}}$, that we will call the bath. Let us assume we have some trial wave function for the whole system, and Schmidt decompose it as $|\Phi\rangle= \sum_i \lambda_i |\alpha_i \rangle |\beta_i \rangle$, where the $|\alpha_i \rangle$ and $|\beta_i \rangle$ are orthogonalized states living on the fragment and the bath, respectively. They define the embedding basis, onto which the original Hamiltonian $\mathcal{H}$ is projected to obtain what a so-called impurity Hamiltonian, $\mathcal{H}_{\mathrm{imp}}=\mathcal{P}_{e}\mathcal{H}\mathcal{P}_{e}$, where $\mathcal{P}_e= \sum_{ij}|\alpha_i \beta_j \rangle \langle \alpha_i \beta_j|$. The motivation of this scheme is that if $|\Phi\rangle$ is the true ground state of ${\cal H}$, then by construction both Hamiltonians share the same ground state. The ground state $|\Phi_{\mathrm{imp}}\rangle$ of $\mathcal{H}_{\mathrm{imp}}$ can therefore be used to compute expectation values of $\mathcal{H}$. The crucial point is that the size of the embedding basis is determined by the size of the fragment. 
As a consequence, $\mathcal{H}_{\mathrm{imp}}$ has a much smaller Hilbert space than $\mathcal{H}$, and its ground state can be obtained using efficient numerical solvers. 

Since the exact solution is not known, the fundamental approximation of DET is to build the embedding basis from $|\Phi\rangle$ taken as a Slater-type wave function~\cite{Knizia2012}. This considerably simplifies the computational treatment, as the Schmidt fragment and bath states are then written in terms of single-particle states~\cite{Klich2006,Peschel2012}. Thus, the impurity Hamiltonian construction becomes a simple change of a single-particle basis, instead of a generically non-trivial many-body projection. 
In practice, $|\Phi\rangle$ is chosen to be the ground state of a certain effective one-body Hamiltonian $h$. This allows tuning the embedding basis by self-consistently updating $h$ such that the one-body density matrices given by $|\Phi\rangle$ and $|\Phi_{\mathrm{imp}}\rangle$ match~\cite{Knizia2012}. In the original introduction of DMET, the matching condition is that the two matrices are entirely identical. 
~\footnote{This condition is ill-defined, as the one-body density matrix of a Slater determinant is idempotent, a property generally not ve%ified by the one obtained from a correlated wave function like \unexpanded{$|\Phi_{\mathrm{imp}}\rangle$}, 
see Ref.~\protect\onlinecite{Bulik2014a}. 
} 
DET is a variant in which this condition is relaxed and imposed only for their diagonals in the impurity block~\cite{Bulik2014a}.

The DMET has been first proposed and tested on 1D and 2D Hubbard models, showing it reproduces accurately the ground state energies~\cite{Knizia2012}, and locates the Mott metal-insulator transition is with comparable efficiency to DMFT. 
It has then been applied to various Hubbard models~\cite{Chen2014,Leblanc2015,Zheng2016}. Several variations and extensions have been proposed, such as DET~\cite{Bulik2014a}, formulations to restore translational invariance~\cite{Zheng2017}, for dynamical properties~\cite{Booth2015}, electron-phonon models~\cite{Sandhoefer2016,Reinhard2018}, Anderson impurity models~\cite{Mukherjee2017}, non-equilibrium electron dynamics~\cite{Kretchmer2018}, and a different scheme has been designed for spin models~\cite{Fan2015,Klaas2017}. DMET has also been applied to a variety of quantum chemistry problems~\cite{Knizia2013,Bulik2014b,Tsuchimochi2015,Wouters2016,Yamazaki2018,Ye2018,Pham2018}.

However, the quality of the wave function $|\Phi_{\mathrm{imp}}\rangle$ has essentially been tested by calculating the ground state energy~\cite{Knizia2012,Bulik2014a,Zheng2017}, which is not an enlightening quantity in general. 
Here, in particular, its accuracy does not guarantee the efficiency of the method: since DMET is not variational~\cite{Knizia2012}, it can deliver energies lower than the exact ones~\cite{Bulik2014a}. Also, in the context of strongly correlated electronic models, only on-site interaction has been included, therefore the performance of the method in the presence of inter-site interactions needs to be studied. In this paper, we apply DET to spinless fermion models. Benchmarking it on a reference chain model, we find that the wave function produced reproduces not only local fragment observables such as the energy or the correlations with high accuracy but also the entanglement content. 
We accordingly show that DET can detect phase transitions, and is applicable in 2D systems. 

The generic form of the $t$-$V$ Hamiltonian we deal with reads
\begin{equation}
{\cal H}= \sum_{ij} \big( t_{ij} c_i^\dagger c_j + {\rm H.c}\big) + \sum_{ij} V_{ij} n_i n_j, 
\label{eq:ham_tV}
\end{equation}
where $c_i^\dagger$ ($c_i$) is the creation (anihilation) operator of spinless fermions, $n_i=c_i^\dagger c_i$. The embedding basis is constructed from the ground state of the one-body Hamiltonian
\begin{equation}
h = \sum_{\langle ij\rangle} \big( t_{ij} c_i^\dagger c_j + {\rm H.c}) + \sum_{i} u_{i} n_{i},
\end{equation}
where we have introduced the DET effective one-body potentials $u_{i}$, $i=1,...,N_{\mathrm{imp}}$, whose purpose is to adapt the embedding basis. It is defined on the fragment and periodically repeated over the lattice. This choice of taking $h$ as the one-body part of $\mathcal{H}$ is the simplest one, but more sophisticated alternatives are possible, such as doing a mean-field calculation.~\cite{Bulik2014a,Wouters2016} The ground state $|\Phi\rangle$ of $h$ is then easily Schmidt decomposed,~\cite{Klich2006} and we obtain a set of $N_{\mathrm{imp}}$ single-particles states $f_{\alpha}$ for the fragment, and similarly $N_{\mathrm{imp}}$ states $b_{\alpha}$ for the bath.~\footnote{We assume that the number of fermions is larger that \unexpanded{$N_{\mathrm{imp}}$}, see Ref.~\onlinecite{Klich2006}.} The dimension of the single-particle embedding basis is thus $2N_{\mathrm{imp}}$, considerably smaller than of the original problem.

The impurity Hamiltonian is then expressed as
\begin{equation}
\mathcal{H}_{\mathrm{imp}} = \sum_{\alpha\beta} \tilde t_{\alpha\beta} e_\alpha^\dagger e_\beta 
+ \sum_{\alpha\beta\gamma\delta} \tilde V_{\alpha\beta\gamma\delta} e_\alpha^\dagger e_\beta  e_\gamma^\dagger e_\delta,
\label{himp;tv}
\end{equation}
where the $2N_{\mathrm{imp}}$ creation operators $e_{\alpha}^{\dagger}$ collectively denotes the fragment and bath states. Several comments are in order here. First, we remark that the fragment states can chosen to be the original ones, $f^{\dagger}_{\alpha}=c^{\dagger}_{\alpha}$,~\cite{Knizia2013,Bulik2014a} such that the intra fragment part of Eq.~\ref{himp;tv} is the original Hamiltonian. We also point out that the Hamiltonian (\ref{himp;tv}) is generically much more complex than the original one, namely in practice it contains assisted and pair hoppings. Third, it is worth noting that the embedding yields a finite bath, thus no bath scaling is needed, contrary to DMFT. To find the ground state $|\Phi\rangle_{\mathrm{imp}}$, we employ a standard exact diagonalization approach, which is possible for small enough fragments. Following the DET prescription,~\cite{Bulik2014a} the effective potentials $u_{i}$ are updated to minimize the difference $| \langle f^{\dagger}_{\alpha} f_{\alpha} \rangle_{\Phi_{\mathrm{imp}}} - \langle f^{\dagger}_{\alpha} f_{\alpha} \rangle_{\Phi} |^{2}$. Physically, this condition ensures that the fragment in the impurity model has the correct average number of particles if the minimum is zero, and the meaning of the effective potential becomes clear. In practical calculations, the perfect match between the two quantities is not always achieved, but we have verified that deviations of the fragment density remain small, of order $10^{-3}$ or less.

We start by presenting the results for the $t_1$-$V_1$-$V_2$ chain at half-filling. We set the nearest-neighbours hopping term $t_{ij}=t_{1}=1$ and the next nearest-neighbour $t_2=0$ and the corresponding interactions are $V_{ij}=V_1$ and $V_2$, respectively. The case $V_2=0$ can be mapped to the exactly solvable spin-1/2 XXZ chain, where a Berezinskii-Kosterlitz-Thouless (BKT) transition from a gapless Tomonaga-Luttinger liquid to the gapped Ising phase takes place at the Heisenberg point. In the fermionic language, these are metallic and charge density wave (CDW) phases, respectively, at $V_1=2$. Introducing a finite $V_2$, the model no longer has an exact solution and the transition point is shifted to larger $V_1$ values. Since the model is one-dimensional, we can use the efficient density matrix renormalization group~\cite{White1992} algorithm (DMRG) to assess the correctness of the DET results.

%*%*%*%*%*%*%*%*%*%*%*%*%*%*%*%*%*%*%*
%  fig1
%*%*%*%*%*%*%*%*%*%*%*%*%*%*%*%*%*%*%*
\begin{figure}[tbp]
\begin{center}
\includegraphics[width=\columnwidth,clip]{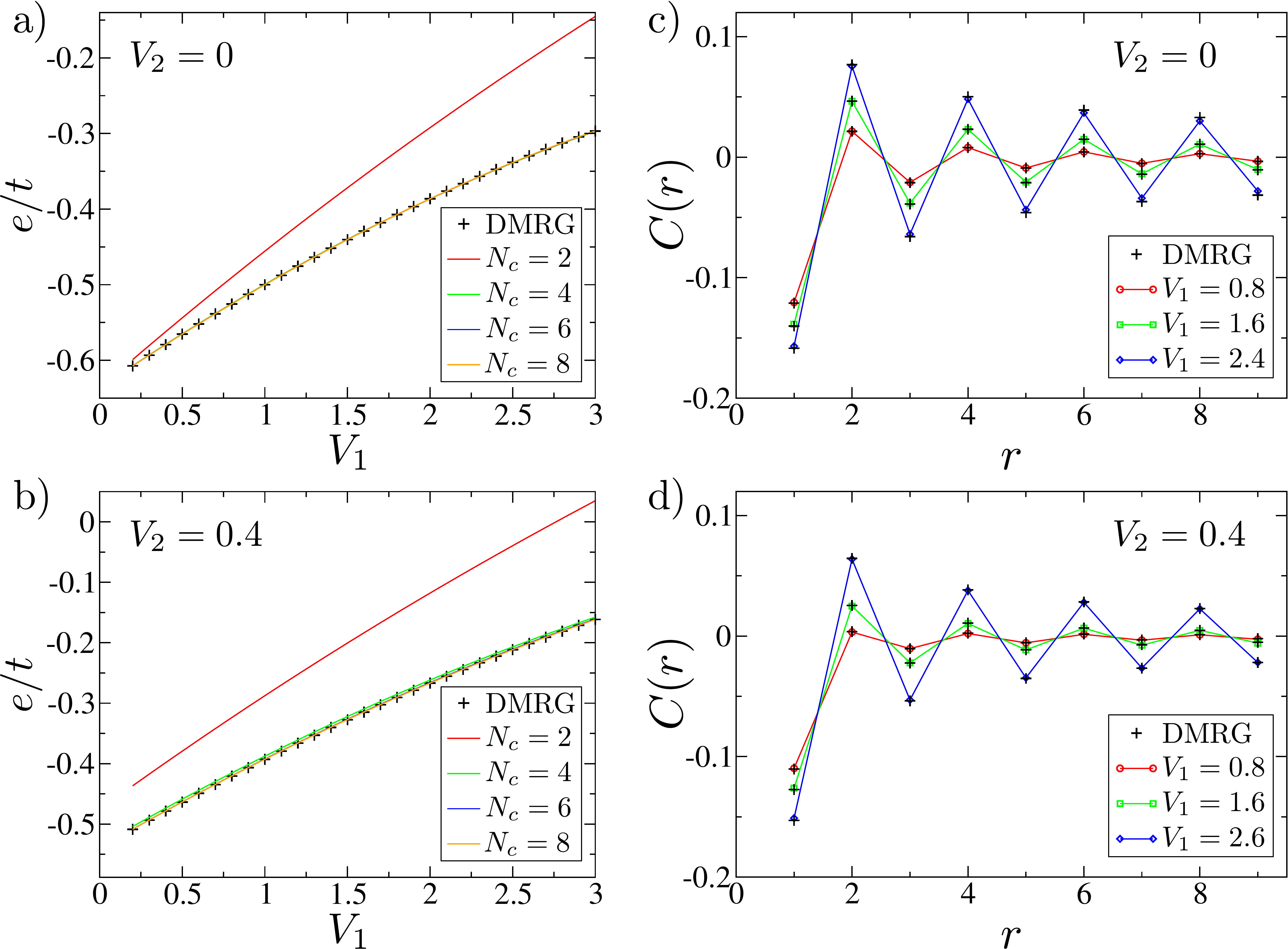}
\end{center}
\caption{(Color online) a) and b) Comparison between DMRG and DET results for the energy per site of the $t_1$-$V_1$-$V_2$ chain at half-filling as a function of $V_1$, for various fragment sizes $N_{\mathrm{imp}}$ and a) $V_2 = 0$, b) $V_2 = 0.4$. c) and d) Connected density correlations $C(r) = \langle n_{i} n_{i+r} \rangle - \langle n_{i} \rangle \langle n_{i+r} \rangle$ obtained by DET for $N_{\mathrm{imp}}=10$ and various values of $V_{1}$. Cross symbols are the DMRG results with $N=128$. 
}
\label{fig:chain_1}
\end{figure}
%*%*%*%*%*%*%*%*%*%*%*%*%*%*%*%*%*%*%*
%*%*%*%*%*%*%*%*%*%*%*%*%*%*%*%*%*%*%*

Figure~\ref{fig:chain_1} a) and \ref{fig:chain_1} b) show the energy per site for $V_{2}=0$ and $0.4$, measured at the center of the fragment to reduce boundary effects, computed with DET for various $N_{\mathrm{imp}}$ with $N=72N_{\mathrm{imp}}$, and compared to the DMRG values with $N=128$ sites. We observe that the smallest $N_{\mathrm{imp}}=2$ gives very poor results, even at small $V_1$, whereas the next size $N_{\mathrm{imp}}=4$ already provides good accuracy. For larger sizes, the deviations from DMRG become negligible. The reason is that for $N_{\mathrm{imp}}=2$ the whole fragment interacts with the bath in Eq.(3), which translates into strong boundary effects. This does not happen in the Hubbard case, where a single site is sufficient to obtain accurate energy.~\cite{Knizia2012}

We report in Figs.~\ref{fig:chain_1}c) and \ref{fig:chain_1}d) the density correlations $C(r) = \langle n_{i} n_{i+r} \rangle - \langle n_{i} \rangle \langle n_{i+r} \rangle$, with the reference site $i$ taken on the edge of a $N_{\mathrm{imp}}=10$ sites fragment, and the correlations computed along the whole fragment. While it would seem more natural to choose the reference site at the center of the fragment, we have verified that both setups give very similar results, but the present one allows us to reach larger $r$. Remarkably, we obtain almost coincident values to that of DMRG with $N=128$, even for the largest $r$ where deviations due to the fragment boundary could be expected~\footnote{Using the relations between the impurity/bath and the original fermions, correlations for larger $r$ could in principle be evaluated. However, they pick up a contribution from the approximate wave function \unexpanded{$|\Phi\rangle$} in addition to the one from \unexpanded{$|\Phi_{\mathrm{imp}}\rangle$}. In our case, we have verified that they immediately deviate from the correct values, but this is expected since we use the simplest possible approximate \unexpanded{$\Phi\rangle$.} when constructing the impurity Hamiltonian.}. This demonstrates that DET can reproduce the correlations obtained for a very large system. In particular, this is true in the gapped CDW phase ($V_1>2$ for $V_2$=0), even if the $|\Phi \rangle$ used does not have the CDW pattern and is thus qualitatively different from $|\Phi_{\mathrm{imp}} \rangle$. An important remark is that we have verified that using an \textit{odd} $N_{\mathrm{imp}}$ does not alter the results, despite the fragment not being commensurate with the CDW state. 
This is a striking difference with DMFT, where the choice of the shape and size of the cluster severely influences the result, 
where particularly the incompatibilities between the odd/even cluster choice and the symmetry breaking pattern matters. 

The quality of the correlated wave function $|\Phi_{\mathrm{imp}}\rangle$ can be more precisely tested by examining the entanglement between the fragment and the bath, a non-local quantity. To measure entanglement properties, we divide the impurity model into two subsystems $A$ and $B$, where $A$ contains $N_A$ sites in the fragment and $B$ the remaining $N_B = 2N_{\mathrm{imp}} - N_A$ sites. From the eigenvalues $\lambda_k$ of the reduced density matrix of subsystem $A$, $\rho_{A}=\mathrm{Tr}_{B}(|\Phi_{\mathrm{imp}}\rangle \langle \Phi_{\mathrm{imp}}|)$, the von Neumann entanglement entropy reads $S_{A}=-\mathrm{Tr}(\rho_{A}\mathrm{log}\rho_{A})=-\sum_{k}\lambda_k\mathrm{log}\lambda_k$. Another quantity of interest is the so-called entanglement spectrum~\cite{Li2008} (ES), which corresponds to the eigenvalues $\zeta_k$ of the entanglement Hamiltonian $\mathcal{H}_A$ defined by $\rho_{A} = \mathrm{e}^{-\mathcal{H_A}}$, from which we see that $\zeta_k = -\mathrm{log}\lambda_k$. %Using the charge conservation, we can block-diagonalize $\rho_A$ in sectors labeled by the charge number $n_A$ to further classify the $\lambda_k$.

\begin{figure}[t]
\begin{center}
\includegraphics[width=\columnwidth,clip]{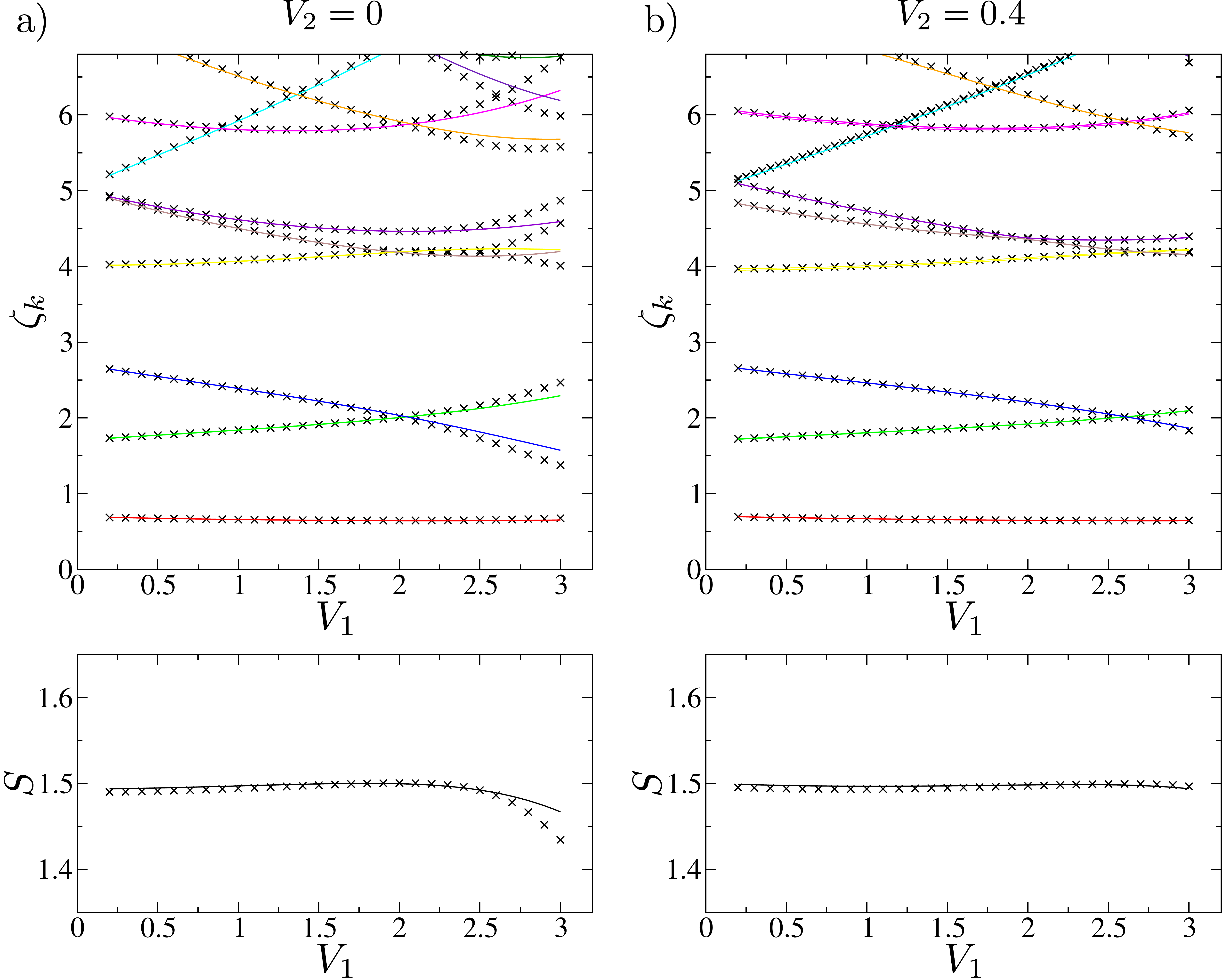}
\end{center}
\caption{(Color online) Top panel: entanglement spectrum computed in DET between the fragment $N_{\mathrm{imp}}=10$ and the bath, $N=720$ and for a) $V_2 = 0$ and b) $V_2 = 0.4$. Bottom panel: von Neumann entanglement entropy for the same parameter values. The cross symbols are the reference values from DMRG with periodic boundary conditions and $N=64$.}
\label{fig:chain_2}
\end{figure}

We plot in the top panel of Fig.~\ref{fig:chain_2} the lowest levels of the ES as a function of $V_1$, for $V_2 = 0$ and $0.4$, together with the DMRG values for $N=64$. The calculations are done for $N_{\mathrm{imp}}=10$ and $N_{A}=N_{\mathrm{imp}}$~\footnote{Comparable results are obtained when $A$ only contains a subset of the fragment}. We see that the lowest level is very well reproduced for all values of $V_1$, in both the metallic and the CDW phases. For higher levels, the agreement is good in all the metallic phase and up to the transition point, where the accuracy starts to deteriorate. Nonetheless, the trends of the various levels are still in fairly good agreement on the CDW side. The entanglement entropy, which is essentially the ES integrated and dominated by the lowest levels, is close to the DMRG values (bottom panel of Fig.~\ref{fig:chain_2}).

%*%*%*%*%*%*%*%*%*%*%*%*%*%*%*%*%*%*%*
%  fig2
%*%*%*%*%*%*%*%*%*%*%*%*%*%*%*%*%*%*%*
%*%*%*%*%*%*%*%*%*%*%*%*%*%*%*%*%*%*%*
%*%*%*%*%*%*%*%*%*%*%*%*%*%*%*%*%*%*%*

In our case, the BKT transition between the metal and the CDW phases is revealed by changes in the ES, namely by a level crossing~\footnote{Notice that in the case $V_2 = 0$, the crossing at $V_1 = 2$ is dictated by the extra degeneracies coming from the SU(2) symmetry at this point, which coincides with the transition point. However, the DET method does not conserve this symmetry when constructing the impurity Hamiltonian, therefore the crossing we observe is not constrained, and in fact, is slightly shifted to larger $V_1\geq 2$.} 
with a change of degeneracy of the first excited level, reminiscent of what happens in the real energy spectrum~\cite{Kitazawa1997}. 
In the gapless side, it is twofold degenerate in the sectors $n_A = N_A/2 \pm 1$, while in the gapped phase it is unique and belongs to the sector $n_A=N_A/2$. Notice that there is no anomaly in the energy in Fig.~\ref{fig:chain_1} at the transition point.

\begin{figure}[t]
\begin{center}
\includegraphics[width=\columnwidth,clip]{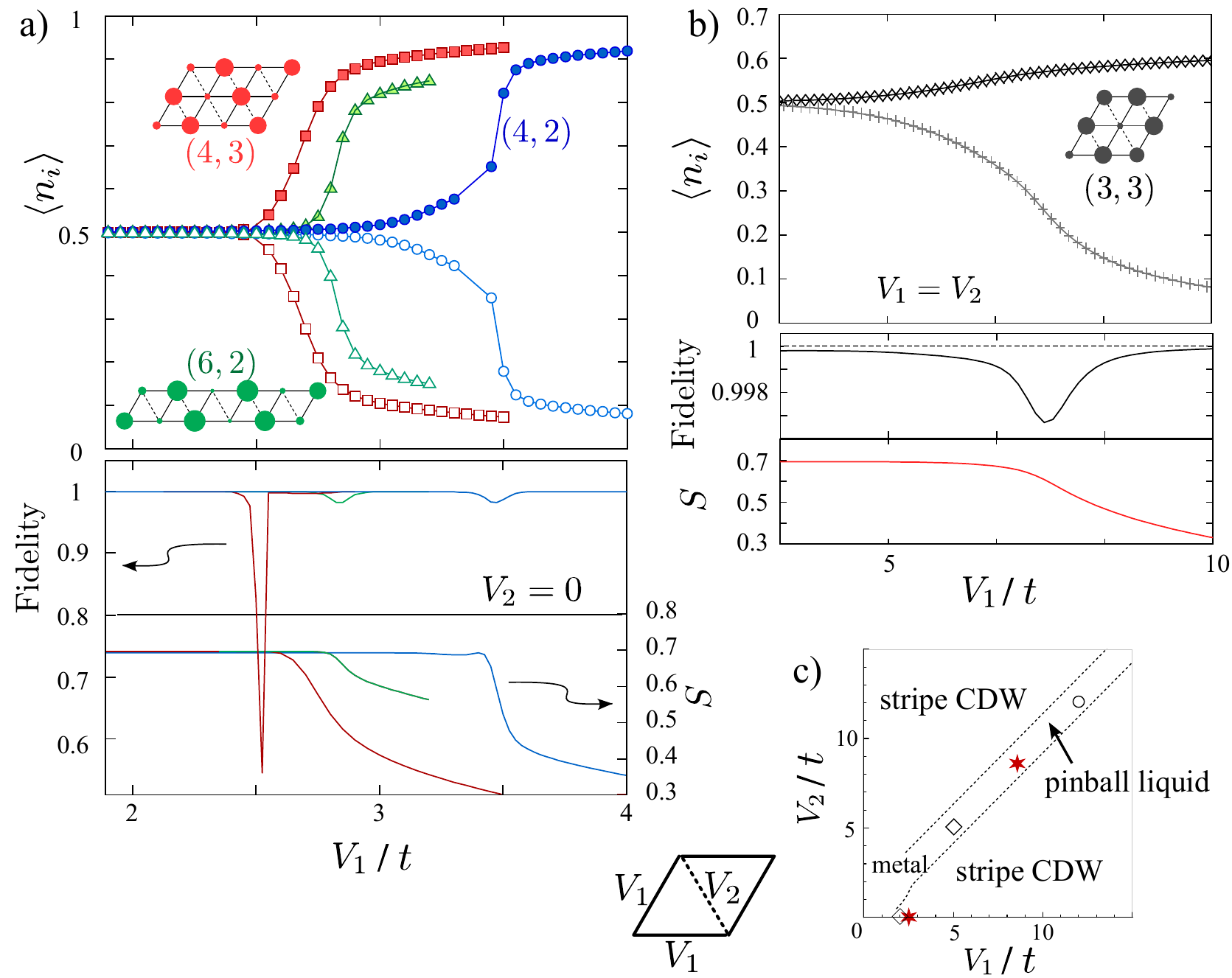}
\end{center}
\caption{(Color online) DET results for the $t$-$V$ model on the anisotropic triangular lattice with two different nearest neighbour interactions $V_1$ and $V_2$ bonds, and uniform hoppings $t_1=t_2=t$. a) Metal-CDW transition along the $V_2=0$ axis revealed by the fragment particle densities $\langle n_i \rangle$, the fidelity and the von Neumann entanglement entropy $S$. b) Metal-pinball liquid transition along the $V_1=V_2$ line shown for the same quantities. c) Phase diagram sketch as function of $V_1$ and $V_2$ from DMRG calculations~\cite{Nishimoto2009}. Red star symbols indicate the metal-CDW and metal-pinball liquid transition points from the DET results of a) and b). Diamond and circle symbols stand for the DMRG and VMC estimates, respectively.}
\label{fig:triangular}
\end{figure}

To further demonstrate the efficiency of DET, we move to a 2D $t$-$V$ Hamiltonian on the anisotropic triangular lattice at half-filling. We take the interactions along two of the three bond directions as $V_1$ and $V_2$ for the remaining one, as illustrated in Fig.~\ref{fig:triangular}c), and the nearest-neighbor hopping $t=1$ as uniform. The phase diagram at half-filling has been numerically studied by DMRG~\cite{Nishimoto2009} (see Fig.~\ref{fig:triangular}c)), exact diagonalization~\cite{Hotta2006a,Hotta2006b} and variational Monte-Carlo (VMC) method~\cite{Miyazaki2009}. Although the overall features of the phase diagram are consistent among them, the quantitative location of some transitions still has large uncertainties.

%*%*%*%*%*%*%*%*%*%*%*%*%*%*%*%*%*%*%*
%  fig3
%*%*%*%*%*%*%*%*%*%*%*%*%*%*%*%*%*%*%*
%*%*%*%*%*%*%*%*%*%*%*%*%*%*%*%*%*%*%*
%*%*%*%*%*%*%*%*%*%*%*%*%*%*%*%*%*%*%*

In Fig.~\ref{fig:triangular}a), we plot the fragment charge densities $\langle n_i \rangle$, the fidelity $|\langle \Psi_{\rm imp}(V_1)|\Psi_{\rm imp}(V_1+dV)\rangle|$ and the entanglement entropy $S$ for $N_A=N_{\mathrm{imp}}$ across the metal-stripe CDW transition, as functions of $V_1$ for $V_2=0$, where the transition takes place around $V_{1;c}\sim 2$ in DMRG. The fragment is labelled by its sides lengths $(\ell_{x},\ell_{y})$ (see Fig.~\ref{fig:triangular}), with $N_{\mathrm{imp}}=\ell_{x}\ell_{y}$, and we perform calculations for $(4,2), (6,2)$ and $(4,3)$ fragments in a $N=60\times 60$ system. The transition out of the metallic phase is signaled by the appearance of high and low charge densities respecting the stripe order. The transition point is more accurately evaluated from the dip in the fidelity, which is also consistent with the onset of the reduction of $S$. It has a clear fragment size dependence, but we find that the values $V_{1,c}=3.47, 2.85$ and $2.52$ for $(4,2), (6,2)$ and $(4,3)$, respectively, quickly converges with larger fragment size and more isotropic shape towards a value definitevly larger than the DMRG estimate~\footnote{Since DMRG favours low entangelement states, it may overestimate the stability of the CDW phase}. Also, we see from the $V_1$-dependence of $\langle n_i \rangle$ and $S$, that $(4,2)$-ones approach the value of the extrapolated $(4,3)$, and $(6,2)$-ones does so to $(4,2)$, thus the different fragment calculations are fully consistent with each other. Therefore, we suppose that the $(4,3)$ results are reasonably good, and the true transition point should be $V_{1;c}=2-2.5$.

This model also has a more complicated transition along the $V_1=V_2$ line, from the metal to a "pinball liquid" at large interaction. In this phase, a partial charge order coexists with a metallic behavior: in one of the three sublattices, holes of fermions produce a symmetry breaking long range order, and the rest of the sites form a honeycomb metal. In previous methods it has been difficult to locate the phase boundary, with DMRG and VMC finding very different $V_{1;c}\simeq 6$ and $\simeq$ 12, respectively, even though these were obtained from a careful finite-size scaling analysis. We demonstrate that DET can detect the transition. As plotted in Fig.~\ref{fig:triangular}b) for a $(3,3)$ fragment, the fidelity exhibits a dip around $V_{1;c}=8.6$, where the entanglement entropy also starts to decrease. The charge densities show the development of a pinball-like charge distribution. The transition point is found in between the two methods mentioned above, which should be reasonable enough. Interestingly, the charge differentiation observed in $\langle n_i \rangle$ develops rather slowly from $V_1=V_2\sim 0$, in contrast to the case of the metal-CDW transition, and does not indicate any anomaly at $V_{1;c}$. Despite that, the fact that DET produces a wave function still allows us to evaluate the transition point by using the fidelity and the entanglement related quantities.

To summarize, we have shown that DET can be a performant method to study spinless fermion models. 
It is an embedding type method which eventually yields a wave function allowing the use of fidelity and entanglement properties as markers for phase transitions. The standard physical quantities such as energy and two-point correlators keep their accuracy independent of 
the shape and size of the impurity fragment, 
indicating that it should be differentiated from the other cluster methods like DMET. 
\\
{\it Acknowledgements.} 
X.P. was supported by FY2015 JSPS Postdoctoral Fellowship for North American and European Researchers, 
and RIKEN iTHES Project. He also acknowledge T. Momoi and A. Furusaki for many supports. 
C.H. was supported by JSPS KAKENHI Grants No. JP17K05533, No. JP18H01173, and No. JP17K05497.

%%%%%%%%%%%%%%%%%%%%%%%%%%%%%%%%%%%%%%%%%%%%%%%%%%%%%%
%%%%%%%%%%%%%%%%%%%% BIBLIOGRAPHY %%%%%%%%%%%%%%%%%%%%
%%%%%%%%%%%%%%%%%%%%%%%%%%%%%%%%%%%%%%%%%%%%%%%%%%%%%%

\bibliographystyle{./apsrev4-1}
\bibliography{dmet}

\begin{thebibliography}{46}%
\makeatletter
\providecommand \@ifxundefined [1]{%
 \@ifx{#1\undefined}
}%
\providecommand \@ifnum [1]{%
 \ifnum #1\expandafter \@firstoftwo
 \else \expandafter \@secondoftwo
 \fi
}%
\providecommand \@ifx [1]{%
 \ifx #1\expandafter \@firstoftwo
 \else \expandafter \@secondoftwo
 \fi
}%
\providecommand \natexlab [1]{#1}%
\providecommand \enquote  [1]{``#1''}%
\providecommand \bibnamefont  [1]{#1}%
\providecommand \bibfnamefont [1]{#1}%
\providecommand \citenamefont [1]{#1}%
\providecommand \href@noop [0]{\@secondoftwo}%
\providecommand \href [0]{\begingroup \@sanitize@url \@href}%
\providecommand \@href[1]{\@@startlink{#1}\@@href}%
\providecommand \@@href[1]{\endgroup#1\@@endlink}%
\providecommand \@sanitize@url [0]{\catcode `\\12\catcode `\$12\catcode
  `\&12\catcode `\#12\catcode `\^12\catcode `\_12\catcode `\%12\relax}%
\providecommand \@@startlink[1]{}%
\providecommand \@@endlink[0]{}%
\providecommand \url  [0]{\begingroup\@sanitize@url \@url }%
\providecommand \@url [1]{\endgroup\@href {#1}{\urlprefix }}%
\providecommand \urlprefix  [0]{URL }%
\providecommand \Eprint [0]{\href }%
\providecommand \doibase [0]{http://dx.doi.org/}%
\providecommand \selectlanguage [0]{\@gobble}%
\providecommand \bibinfo  [0]{\@secondoftwo}%
\providecommand \bibfield  [0]{\@secondoftwo}%
\providecommand \translation [1]{[#1]}%
\providecommand \BibitemOpen [0]{}%
\providecommand \bibitemStop [0]{}%
\providecommand \bibitemNoStop [0]{.\EOS\space}%
\providecommand \EOS [0]{\spacefactor3000\relax}%
\providecommand \BibitemShut  [1]{\csname bibitem#1\endcsname}%
\let\auto@bib@innerbib\@empty
%</preamble>
\bibitem [{\citenamefont {Bardeen}\ \emph {et~al.}(1957)\citenamefont
  {Bardeen}, \citenamefont {Cooper},\ and\ \citenamefont
  {Schrieffer}}]{BCS1957}%
  \BibitemOpen
  \bibfield  {author} {\bibinfo {author} {\bibfnamefont {J.}~\bibnamefont
  {Bardeen}}, \bibinfo {author} {\bibfnamefont {L.~N.}\ \bibnamefont {Cooper}},
  \ and\ \bibinfo {author} {\bibfnamefont {J.~R.}\ \bibnamefont {Schrieffer}},\
  }\href {\doibase 10.1103/PhysRev.106.162} {\bibfield  {journal} {\bibinfo
  {journal} {Phys. Rev.}\ }\textbf {\bibinfo {volume} {106}},\ \bibinfo {pages}
  {162} (\bibinfo {year} {1957})}\BibitemShut {NoStop}%
\bibitem [{\citenamefont {Laughlin}(1983)}]{Laughlin1983}%
  \BibitemOpen
  \bibfield  {author} {\bibinfo {author} {\bibfnamefont {R.~B.}\ \bibnamefont
  {Laughlin}},\ }\href {\doibase 10.1103/PhysRevLett.50.1395} {\bibfield
  {journal} {\bibinfo  {journal} {Phys. Rev. Lett.}\ }\textbf {\bibinfo
  {volume} {50}},\ \bibinfo {pages} {1395} (\bibinfo {year}
  {1983})}\BibitemShut {NoStop}%
\bibitem [{\citenamefont {Zanardi}\ and\ \citenamefont
  {Paunkovi\ifmmode~\acute{c}\else \'{c}\fi{}}(2006)}]{Zanardi2006}%
  \BibitemOpen
  \bibfield  {author} {\bibinfo {author} {\bibfnamefont {P.}~\bibnamefont
  {Zanardi}}\ and\ \bibinfo {author} {\bibfnamefont {N.}~\bibnamefont
  {Paunkovi\ifmmode~\acute{c}\else \'{c}\fi{}}},\ }\href {\doibase
  10.1103/PhysRevE.74.031123} {\bibfield  {journal} {\bibinfo  {journal} {Phys.
  Rev. E}\ }\textbf {\bibinfo {volume} {74}},\ \bibinfo {pages} {031123}
  (\bibinfo {year} {2006})}\BibitemShut {NoStop}%
\bibitem [{\citenamefont {Cozzini}\ \emph {et~al.}(2007)\citenamefont
  {Cozzini}, \citenamefont {Ionicioiu},\ and\ \citenamefont
  {Zanardi}}]{Cozzini2007}%
  \BibitemOpen
  \bibfield  {author} {\bibinfo {author} {\bibfnamefont {M.}~\bibnamefont
  {Cozzini}}, \bibinfo {author} {\bibfnamefont {R.}~\bibnamefont {Ionicioiu}},
  \ and\ \bibinfo {author} {\bibfnamefont {P.}~\bibnamefont {Zanardi}},\ }\href
  {\doibase 10.1103/PhysRevB.76.104420} {\bibfield  {journal} {\bibinfo
  {journal} {Phys. Rev. B}\ }\textbf {\bibinfo {volume} {76}},\ \bibinfo
  {pages} {104420} (\bibinfo {year} {2007})}\BibitemShut {NoStop}%
\bibitem [{\citenamefont {Albuquerque}\ \emph {et~al.}(2010)\citenamefont
  {Albuquerque}, \citenamefont {Alet}, \citenamefont {Sire},\ and\
  \citenamefont {Capponi}}]{Albuquerque2010}%
  \BibitemOpen
  \bibfield  {author} {\bibinfo {author} {\bibfnamefont {A.~F.}\ \bibnamefont
  {Albuquerque}}, \bibinfo {author} {\bibfnamefont {F.}~\bibnamefont {Alet}},
  \bibinfo {author} {\bibfnamefont {C.}~\bibnamefont {Sire}}, \ and\ \bibinfo
  {author} {\bibfnamefont {S.}~\bibnamefont {Capponi}},\ }\href {\doibase
  10.1103/PhysRevB.81.064418} {\bibfield  {journal} {\bibinfo  {journal} {Phys.
  Rev. B}\ }\textbf {\bibinfo {volume} {81}},\ \bibinfo {pages} {064418}
  (\bibinfo {year} {2010})}\BibitemShut {NoStop}%
\bibitem [{\citenamefont {Gu}(2010)}]{Gu2010}%
  \BibitemOpen
  \bibfield  {author} {\bibinfo {author} {\bibfnamefont {S.-J.}\ \bibnamefont
  {Gu}},\ }\href {\doibase 10.1142/S0217979210056335} {\bibfield  {journal}
  {\bibinfo  {journal} {International Journal of Modern Physics B}\ }\textbf
  {\bibinfo {volume} {24}},\ \bibinfo {pages} {4371} (\bibinfo {year}
  {2010})}\BibitemShut {NoStop}%
\bibitem [{\citenamefont {Amico}\ \emph {et~al.}(2008)\citenamefont {Amico},
  \citenamefont {Fazio}, \citenamefont {Osterloh},\ and\ \citenamefont
  {Vedral}}]{Amico2008}%
  \BibitemOpen
  \bibfield  {author} {\bibinfo {author} {\bibfnamefont {L.}~\bibnamefont
  {Amico}}, \bibinfo {author} {\bibfnamefont {R.}~\bibnamefont {Fazio}},
  \bibinfo {author} {\bibfnamefont {A.}~\bibnamefont {Osterloh}}, \ and\
  \bibinfo {author} {\bibfnamefont {V.}~\bibnamefont {Vedral}},\ }\href
  {\doibase 10.1103/RevModPhys.80.517} {\bibfield  {journal} {\bibinfo
  {journal} {Rev. Mod. Phys.}\ }\textbf {\bibinfo {volume} {80}},\ \bibinfo
  {pages} {517} (\bibinfo {year} {2008})}\BibitemShut {NoStop}%
\bibitem [{\citenamefont {Eisert}\ \emph {et~al.}(2010)\citenamefont {Eisert},
  \citenamefont {Cramer},\ and\ \citenamefont {Plenio}}]{Eisert2010}%
  \BibitemOpen
  \bibfield  {author} {\bibinfo {author} {\bibfnamefont {J.}~\bibnamefont
  {Eisert}}, \bibinfo {author} {\bibfnamefont {M.}~\bibnamefont {Cramer}}, \
  and\ \bibinfo {author} {\bibfnamefont {M.~B.}\ \bibnamefont {Plenio}},\
  }\href {\doibase 10.1103/RevModPhys.82.277} {\bibfield  {journal} {\bibinfo
  {journal} {Rev. Mod. Phys.}\ }\textbf {\bibinfo {volume} {82}},\ \bibinfo
  {pages} {277} (\bibinfo {year} {2010})}\BibitemShut {NoStop}%
\bibitem [{\citenamefont {Legeza}\ and\ \citenamefont
  {S\'olyom}(2006)}]{Legeza2006}%
  \BibitemOpen
  \bibfield  {author} {\bibinfo {author} {\bibfnamefont {O.}~\bibnamefont
  {Legeza}}\ and\ \bibinfo {author} {\bibfnamefont {J.}~\bibnamefont
  {S\'olyom}},\ }\href {\doibase 10.1103/PhysRevLett.96.116401} {\bibfield
  {journal} {\bibinfo  {journal} {Phys. Rev. Lett.}\ }\textbf {\bibinfo
  {volume} {96}},\ \bibinfo {pages} {116401} (\bibinfo {year}
  {2006})}\BibitemShut {NoStop}%
\bibitem [{\citenamefont {Chepiga}\ \emph {et~al.}(2016)\citenamefont
  {Chepiga}, \citenamefont {Affleck},\ and\ \citenamefont
  {Mila}}]{Chepiga2016}%
  \BibitemOpen
  \bibfield  {author} {\bibinfo {author} {\bibfnamefont {N.}~\bibnamefont
  {Chepiga}}, \bibinfo {author} {\bibfnamefont {I.}~\bibnamefont {Affleck}}, \
  and\ \bibinfo {author} {\bibfnamefont {F.}~\bibnamefont {Mila}},\ }\href
  {\doibase 10.1103/PhysRevB.94.205112} {\bibfield  {journal} {\bibinfo
  {journal} {Phys. Rev. B}\ }\textbf {\bibinfo {volume} {94}},\ \bibinfo
  {pages} {205112} (\bibinfo {year} {2016})}\BibitemShut {NoStop}%
\bibitem [{\citenamefont {Hatsugai}(2006)}]{Hatsugai2006}%
  \BibitemOpen
  \bibfield  {author} {\bibinfo {author} {\bibfnamefont {Y.}~\bibnamefont
  {Hatsugai}},\ }\href {\doibase 10.1143/JPSJ.75.123601} {\bibfield  {journal}
  {\bibinfo  {journal} {Journal of the Physical Society of Japan}\ }\textbf
  {\bibinfo {volume} {75}},\ \bibinfo {pages} {123601} (\bibinfo {year}
  {2006})}\BibitemShut {NoStop}%
\bibitem [{\citenamefont {Bulik}\ \emph
  {et~al.}(2014{\natexlab{a}})\citenamefont {Bulik}, \citenamefont {Scuseria},\
  and\ \citenamefont {Dukelsky}}]{Bulik2014a}%
  \BibitemOpen
  \bibfield  {author} {\bibinfo {author} {\bibfnamefont {I.~W.}\ \bibnamefont
  {Bulik}}, \bibinfo {author} {\bibfnamefont {G.~E.}\ \bibnamefont {Scuseria}},
  \ and\ \bibinfo {author} {\bibfnamefont {J.}~\bibnamefont {Dukelsky}},\
  }\href {\doibase 10.1103/PhysRevB.89.035140} {\bibfield  {journal} {\bibinfo
  {journal} {Phys. Rev. B}\ }\textbf {\bibinfo {volume} {89}},\ \bibinfo
  {pages} {035140} (\bibinfo {year} {2014}{\natexlab{a}})}\BibitemShut
  {NoStop}%
\bibitem [{\citenamefont {Knizia}\ and\ \citenamefont
  {Chan}(2012)}]{Knizia2012}%
  \BibitemOpen
  \bibfield  {author} {\bibinfo {author} {\bibfnamefont {G.}~\bibnamefont
  {Knizia}}\ and\ \bibinfo {author} {\bibfnamefont {G.~K.-L.}\ \bibnamefont
  {Chan}},\ }\href {\doibase 10.1103/PhysRevLett.109.186404} {\bibfield
  {journal} {\bibinfo  {journal} {Phys. Rev. Lett.}\ }\textbf {\bibinfo
  {volume} {109}},\ \bibinfo {pages} {186404} (\bibinfo {year}
  {2012})}\BibitemShut {NoStop}%
\bibitem [{\citenamefont {Klich}(2006)}]{Klich2006}%
  \BibitemOpen
  \bibfield  {author} {\bibinfo {author} {\bibfnamefont {I.}~\bibnamefont
  {Klich}},\ }\href {http://stacks.iop.org/0305-4470/39/i=4/a=L02} {\bibfield
  {journal} {\bibinfo  {journal} {Journal of Physics A: Mathematical and
  General}\ }\textbf {\bibinfo {volume} {39}},\ \bibinfo {pages} {L85}
  (\bibinfo {year} {2006})}\BibitemShut {NoStop}%
\bibitem [{\citenamefont {Peschel}(2012)}]{Peschel2012}%
  \BibitemOpen
  \bibfield  {author} {\bibinfo {author} {\bibfnamefont {I.}~\bibnamefont
  {Peschel}},\ }\href {\doibase 10.1007/s13538-012-0074-1} {\bibfield
  {journal} {\bibinfo  {journal} {Brazilian Journal of Physics}\ }\textbf
  {\bibinfo {volume} {42}},\ \bibinfo {pages} {267} (\bibinfo {year}
  {2012})}\BibitemShut {NoStop}%
\bibitem [{Note1()}]{Note1}%
  \BibitemOpen
  \bibinfo {note} {This condition is ill-defined, as the one-body density
  matrix of a Slater determinant is idempotent, a property generally not vesee
  Ref.~\protect \onlinecite {Bulik2014a}.}\BibitemShut {Stop}%
\bibitem [{\citenamefont {Chen}\ \emph {et~al.}(2014)\citenamefont {Chen},
  \citenamefont {Booth}, \citenamefont {Sharma}, \citenamefont {Knizia},\ and\
  \citenamefont {Chan}}]{Chen2014}%
  \BibitemOpen
  \bibfield  {author} {\bibinfo {author} {\bibfnamefont {Q.}~\bibnamefont
  {Chen}}, \bibinfo {author} {\bibfnamefont {G.~H.}\ \bibnamefont {Booth}},
  \bibinfo {author} {\bibfnamefont {S.}~\bibnamefont {Sharma}}, \bibinfo
  {author} {\bibfnamefont {G.}~\bibnamefont {Knizia}}, \ and\ \bibinfo {author}
  {\bibfnamefont {G.~K.-L.}\ \bibnamefont {Chan}},\ }\href {\doibase
  10.1103/PhysRevB.89.165134} {\bibfield  {journal} {\bibinfo  {journal} {Phys.
  Rev. B}\ }\textbf {\bibinfo {volume} {89}},\ \bibinfo {pages} {165134}
  (\bibinfo {year} {2014})}\BibitemShut {NoStop}%
\bibitem [{\citenamefont {LeBlanc}\ \emph {et~al.}(2015)\citenamefont
  {LeBlanc}, \citenamefont {Antipov}, \citenamefont {Becca}, \citenamefont
  {Bulik}, \citenamefont {Chan}, \citenamefont {Chung}, \citenamefont {Deng},
  \citenamefont {Ferrero}, \citenamefont {Henderson}, \citenamefont
  {Jim\'enez-Hoyos}, \citenamefont {Kozik}, \citenamefont {Liu}, \citenamefont
  {Millis}, \citenamefont {Prokof'ev}, \citenamefont {Qin}, \citenamefont
  {Scuseria}, \citenamefont {Shi}, \citenamefont {Svistunov}, \citenamefont
  {Tocchio}, \citenamefont {Tupitsyn}, \citenamefont {White}, \citenamefont
  {Zhang}, \citenamefont {Zheng}, \citenamefont {Zhu},\ and\ \citenamefont
  {Gull}}]{Leblanc2015}%
  \BibitemOpen
  \bibfield  {author} {\bibinfo {author} {\bibfnamefont {J.~P.~F.}\
  \bibnamefont {LeBlanc}}, \bibinfo {author} {\bibfnamefont {A.~E.}\
  \bibnamefont {Antipov}}, \bibinfo {author} {\bibfnamefont {F.}~\bibnamefont
  {Becca}}, \bibinfo {author} {\bibfnamefont {I.~W.}\ \bibnamefont {Bulik}},
  \bibinfo {author} {\bibfnamefont {G.~K.-L.}\ \bibnamefont {Chan}}, \bibinfo
  {author} {\bibfnamefont {C.-M.}\ \bibnamefont {Chung}}, \bibinfo {author}
  {\bibfnamefont {Y.}~\bibnamefont {Deng}}, \bibinfo {author} {\bibfnamefont
  {M.}~\bibnamefont {Ferrero}}, \bibinfo {author} {\bibfnamefont {T.~M.}\
  \bibnamefont {Henderson}}, \bibinfo {author} {\bibfnamefont {C.~A.}\
  \bibnamefont {Jim\'enez-Hoyos}}, \bibinfo {author} {\bibfnamefont
  {E.}~\bibnamefont {Kozik}}, \bibinfo {author} {\bibfnamefont {X.-W.}\
  \bibnamefont {Liu}}, \bibinfo {author} {\bibfnamefont {A.~J.}\ \bibnamefont
  {Millis}}, \bibinfo {author} {\bibfnamefont {N.~V.}\ \bibnamefont
  {Prokof'ev}}, \bibinfo {author} {\bibfnamefont {M.}~\bibnamefont {Qin}},
  \bibinfo {author} {\bibfnamefont {G.~E.}\ \bibnamefont {Scuseria}}, \bibinfo
  {author} {\bibfnamefont {H.}~\bibnamefont {Shi}}, \bibinfo {author}
  {\bibfnamefont {B.~V.}\ \bibnamefont {Svistunov}}, \bibinfo {author}
  {\bibfnamefont {L.~F.}\ \bibnamefont {Tocchio}}, \bibinfo {author}
  {\bibfnamefont {I.~S.}\ \bibnamefont {Tupitsyn}}, \bibinfo {author}
  {\bibfnamefont {S.~R.}\ \bibnamefont {White}}, \bibinfo {author}
  {\bibfnamefont {S.}~\bibnamefont {Zhang}}, \bibinfo {author} {\bibfnamefont
  {B.-X.}\ \bibnamefont {Zheng}}, \bibinfo {author} {\bibfnamefont
  {Z.}~\bibnamefont {Zhu}}, \ and\ \bibinfo {author} {\bibfnamefont
  {E.}~\bibnamefont {Gull}} (\bibinfo {collaboration} {Simons Collaboration on
  the Many-Electron Problem}),\ }\href {\doibase 10.1103/PhysRevX.5.041041}
  {\bibfield  {journal} {\bibinfo  {journal} {Phys. Rev. X}\ }\textbf {\bibinfo
  {volume} {5}},\ \bibinfo {pages} {041041} (\bibinfo {year}
  {2015})}\BibitemShut {NoStop}%
\bibitem [{\citenamefont {Zheng}\ and\ \citenamefont {Chan}(2016)}]{Zheng2016}%
  \BibitemOpen
  \bibfield  {author} {\bibinfo {author} {\bibfnamefont {B.-X.}\ \bibnamefont
  {Zheng}}\ and\ \bibinfo {author} {\bibfnamefont {G.~K.-L.}\ \bibnamefont
  {Chan}},\ }\href {\doibase 10.1103/PhysRevB.93.035126} {\bibfield  {journal}
  {\bibinfo  {journal} {Phys. Rev. B}\ }\textbf {\bibinfo {volume} {93}},\
  \bibinfo {pages} {035126} (\bibinfo {year} {2016})}\BibitemShut {NoStop}%
\bibitem [{\citenamefont {Zheng}\ \emph {et~al.}(2017)\citenamefont {Zheng},
  \citenamefont {Kretchmer}, \citenamefont {Shi}, \citenamefont {Zhang},\ and\
  \citenamefont {Chan}}]{Zheng2017}%
  \BibitemOpen
  \bibfield  {author} {\bibinfo {author} {\bibfnamefont {B.-X.}\ \bibnamefont
  {Zheng}}, \bibinfo {author} {\bibfnamefont {J.~S.}\ \bibnamefont
  {Kretchmer}}, \bibinfo {author} {\bibfnamefont {H.}~\bibnamefont {Shi}},
  \bibinfo {author} {\bibfnamefont {S.}~\bibnamefont {Zhang}}, \ and\ \bibinfo
  {author} {\bibfnamefont {G.~K.-L.}\ \bibnamefont {Chan}},\ }\href {\doibase
  10.1103/PhysRevB.95.045103} {\bibfield  {journal} {\bibinfo  {journal} {Phys.
  Rev. B}\ }\textbf {\bibinfo {volume} {95}},\ \bibinfo {pages} {045103}
  (\bibinfo {year} {2017})}\BibitemShut {NoStop}%
\bibitem [{\citenamefont {Booth}\ and\ \citenamefont {Chan}(2015)}]{Booth2015}%
  \BibitemOpen
  \bibfield  {author} {\bibinfo {author} {\bibfnamefont {G.~H.}\ \bibnamefont
  {Booth}}\ and\ \bibinfo {author} {\bibfnamefont {G.~K.-L.}\ \bibnamefont
  {Chan}},\ }\href {\doibase 10.1103/PhysRevB.91.155107} {\bibfield  {journal}
  {\bibinfo  {journal} {Phys. Rev. B}\ }\textbf {\bibinfo {volume} {91}},\
  \bibinfo {pages} {155107} (\bibinfo {year} {2015})}\BibitemShut {NoStop}%
\bibitem [{\citenamefont {Sandhoefer}\ and\ \citenamefont
  {Chan}(2016)}]{Sandhoefer2016}%
  \BibitemOpen
  \bibfield  {author} {\bibinfo {author} {\bibfnamefont {B.}~\bibnamefont
  {Sandhoefer}}\ and\ \bibinfo {author} {\bibfnamefont {G.~K.-L.}\ \bibnamefont
  {Chan}},\ }\href {\doibase 10.1103/PhysRevB.94.085115} {\bibfield  {journal}
  {\bibinfo  {journal} {Phys. Rev. B}\ }\textbf {\bibinfo {volume} {94}},\
  \bibinfo {pages} {085115} (\bibinfo {year} {2016})}\BibitemShut {NoStop}%
\bibitem [{\citenamefont {{Reinhard}}\ \emph {et~al.}(2018)\citenamefont
  {{Reinhard}}, \citenamefont {{Mordovina}}, \citenamefont {{Hubig}},
  \citenamefont {{Kretchmer}}, \citenamefont {{Schollw{\"o}ck}}, \citenamefont
  {{Appel}}, \citenamefont {{Sentef}},\ and\ \citenamefont
  {{Rubio}}}]{Reinhard2018}%
  \BibitemOpen
  \bibfield  {author} {\bibinfo {author} {\bibfnamefont {T.~E.}\ \bibnamefont
  {{Reinhard}}}, \bibinfo {author} {\bibfnamefont {U.}~\bibnamefont
  {{Mordovina}}}, \bibinfo {author} {\bibfnamefont {C.}~\bibnamefont
  {{Hubig}}}, \bibinfo {author} {\bibfnamefont {J.~S.}\ \bibnamefont
  {{Kretchmer}}}, \bibinfo {author} {\bibfnamefont {U.}~\bibnamefont
  {{Schollw{\"o}ck}}}, \bibinfo {author} {\bibfnamefont {H.}~\bibnamefont
  {{Appel}}}, \bibinfo {author} {\bibfnamefont {M.~A.}\ \bibnamefont
  {{Sentef}}}, \ and\ \bibinfo {author} {\bibfnamefont {A.}~\bibnamefont
  {{Rubio}}},\ }\href@noop {} {\bibfield  {journal} {\bibinfo  {journal} {arXiv
  e-prints}\ ,\ \bibinfo {eid} {arXiv:1811.00048}} (\bibinfo {year} {2018})},\
  \Eprint {http://arxiv.org/abs/1811.00048} {arXiv:1811.00048
  [cond-mat.str-el]} \BibitemShut {NoStop}%
\bibitem [{\citenamefont {Mukherjee}\ and\ \citenamefont
  {Reichman}(2017)}]{Mukherjee2017}%
  \BibitemOpen
  \bibfield  {author} {\bibinfo {author} {\bibfnamefont {S.}~\bibnamefont
  {Mukherjee}}\ and\ \bibinfo {author} {\bibfnamefont {D.~R.}\ \bibnamefont
  {Reichman}},\ }\href {\doibase 10.1103/PhysRevB.95.155111} {\bibfield
  {journal} {\bibinfo  {journal} {Phys. Rev. B}\ }\textbf {\bibinfo {volume}
  {95}},\ \bibinfo {pages} {155111} (\bibinfo {year} {2017})}\BibitemShut
  {NoStop}%
\bibitem [{\citenamefont {Kretchmer}\ and\ \citenamefont
  {Chan}(2018)}]{Kretchmer2018}%
  \BibitemOpen
  \bibfield  {author} {\bibinfo {author} {\bibfnamefont {J.~S.}\ \bibnamefont
  {Kretchmer}}\ and\ \bibinfo {author} {\bibfnamefont {G.~K.-L.}\ \bibnamefont
  {Chan}},\ }\href {\doibase 10.1063/1.5012766} {\bibfield  {journal} {\bibinfo
   {journal} {The Journal of Chemical Physics}\ }\textbf {\bibinfo {volume}
  {148}},\ \bibinfo {pages} {054108} (\bibinfo {year} {2018})}\BibitemShut
  {NoStop}%
\bibitem [{\citenamefont {Fan}\ and\ \citenamefont {Jie}(2015)}]{Fan2015}%
  \BibitemOpen
  \bibfield  {author} {\bibinfo {author} {\bibfnamefont {Z.}~\bibnamefont
  {Fan}}\ and\ \bibinfo {author} {\bibfnamefont {Q.-l.}\ \bibnamefont {Jie}},\
  }\href {\doibase 10.1103/PhysRevB.91.195118} {\bibfield  {journal} {\bibinfo
  {journal} {Phys. Rev. B}\ }\textbf {\bibinfo {volume} {91}},\ \bibinfo
  {pages} {195118} (\bibinfo {year} {2015})}\BibitemShut {NoStop}%
\bibitem [{\citenamefont {Gunst}\ \emph {et~al.}(2017)\citenamefont {Gunst},
  \citenamefont {Wouters}, \citenamefont {De~Baerdemacker},\ and\ \citenamefont
  {Van~Neck}}]{Klaas2017}%
  \BibitemOpen
  \bibfield  {author} {\bibinfo {author} {\bibfnamefont {K.}~\bibnamefont
  {Gunst}}, \bibinfo {author} {\bibfnamefont {S.}~\bibnamefont {Wouters}},
  \bibinfo {author} {\bibfnamefont {S.}~\bibnamefont {De~Baerdemacker}}, \ and\
  \bibinfo {author} {\bibfnamefont {D.}~\bibnamefont {Van~Neck}},\ }\href
  {\doibase 10.1103/PhysRevB.95.195127} {\bibfield  {journal} {\bibinfo
  {journal} {Phys. Rev. B}\ }\textbf {\bibinfo {volume} {95}},\ \bibinfo
  {pages} {195127} (\bibinfo {year} {2017})}\BibitemShut {NoStop}%
\bibitem [{\citenamefont {Knizia}\ and\ \citenamefont
  {Chan}(2013)}]{Knizia2013}%
  \BibitemOpen
  \bibfield  {author} {\bibinfo {author} {\bibfnamefont {G.}~\bibnamefont
  {Knizia}}\ and\ \bibinfo {author} {\bibfnamefont {G.~K.-L.}\ \bibnamefont
  {Chan}},\ }\href {\doibase 10.1021/ct301044e} {\bibfield  {journal} {\bibinfo
   {journal} {Journal of Chemical Theory and Computation}\ }\textbf {\bibinfo
  {volume} {9}},\ \bibinfo {pages} {1428} (\bibinfo {year} {2013})},\ \bibinfo
  {note} {pMID: 26587604}\BibitemShut {NoStop}%
\bibitem [{\citenamefont {Bulik}\ \emph
  {et~al.}(2014{\natexlab{b}})\citenamefont {Bulik}, \citenamefont {Chen},\
  and\ \citenamefont {Scuseria}}]{Bulik2014b}%
  \BibitemOpen
  \bibfield  {author} {\bibinfo {author} {\bibfnamefont {I.~W.}\ \bibnamefont
  {Bulik}}, \bibinfo {author} {\bibfnamefont {W.}~\bibnamefont {Chen}}, \ and\
  \bibinfo {author} {\bibfnamefont {G.~E.}\ \bibnamefont {Scuseria}},\ }\href
  {\doibase 10.1063/1.4891861} {\bibfield  {journal} {\bibinfo  {journal} {The
  Journal of Chemical Physics}\ }\textbf {\bibinfo {volume} {141}},\ \bibinfo
  {pages} {054113} (\bibinfo {year} {2014}{\natexlab{b}})}\BibitemShut
  {NoStop}%
\bibitem [{\citenamefont {Tsuchimochi}\ \emph {et~al.}(2015)\citenamefont
  {Tsuchimochi}, \citenamefont {Welborn},\ and\ \citenamefont
  {Van~Voorhis}}]{Tsuchimochi2015}%
  \BibitemOpen
  \bibfield  {author} {\bibinfo {author} {\bibfnamefont {T.}~\bibnamefont
  {Tsuchimochi}}, \bibinfo {author} {\bibfnamefont {M.}~\bibnamefont
  {Welborn}}, \ and\ \bibinfo {author} {\bibfnamefont {T.}~\bibnamefont
  {Van~Voorhis}},\ }\href {\doibase 10.1063/1.4926650} {\bibfield  {journal}
  {\bibinfo  {journal} {The Journal of Chemical Physics}\ }\textbf {\bibinfo
  {volume} {143}},\ \bibinfo {pages} {024107} (\bibinfo {year}
  {2015})}\BibitemShut {NoStop}%
\bibitem [{\citenamefont {Wouters}\ \emph {et~al.}(2016)\citenamefont
  {Wouters}, \citenamefont {Jiménez-Hoyos}, \citenamefont {Sun},\ and\
  \citenamefont {Chan}}]{Wouters2016}%
  \BibitemOpen
  \bibfield  {author} {\bibinfo {author} {\bibfnamefont {S.}~\bibnamefont
  {Wouters}}, \bibinfo {author} {\bibfnamefont {C.~A.}\ \bibnamefont
  {Jiménez-Hoyos}}, \bibinfo {author} {\bibfnamefont {Q.}~\bibnamefont {Sun}},
  \ and\ \bibinfo {author} {\bibfnamefont {G.~K.-L.}\ \bibnamefont {Chan}},\
  }\href {\doibase 10.1021/acs.jctc.6b00316} {\bibfield  {journal} {\bibinfo
  {journal} {Journal of Chemical Theory and Computation}\ }\textbf {\bibinfo
  {volume} {12}},\ \bibinfo {pages} {2706} (\bibinfo {year}
  {2016})}\BibitemShut {NoStop}%
\bibitem [{\citenamefont {{Yamazaki}}\ \emph {et~al.}(2018)\citenamefont
  {{Yamazaki}}, \citenamefont {{Matsuura}}, \citenamefont {{Narimani}},
  \citenamefont {{Saidmuradov}},\ and\ \citenamefont
  {{Zaribafiyan}}}]{Yamazaki2018}%
  \BibitemOpen
  \bibfield  {author} {\bibinfo {author} {\bibfnamefont {T.}~\bibnamefont
  {{Yamazaki}}}, \bibinfo {author} {\bibfnamefont {S.}~\bibnamefont
  {{Matsuura}}}, \bibinfo {author} {\bibfnamefont {A.}~\bibnamefont
  {{Narimani}}}, \bibinfo {author} {\bibfnamefont {A.}~\bibnamefont
  {{Saidmuradov}}}, \ and\ \bibinfo {author} {\bibfnamefont {A.}~\bibnamefont
  {{Zaribafiyan}}},\ }\href@noop {} {\bibfield  {journal} {\bibinfo  {journal}
  {arXiv e-prints}\ ,\ \bibinfo {eid} {arXiv:1806.01305}} (\bibinfo {year}
  {2018})},\ \Eprint {http://arxiv.org/abs/1806.01305} {arXiv:1806.01305
  [quant-ph]} \BibitemShut {NoStop}%
\bibitem [{\citenamefont {Ye}\ \emph {et~al.}(2018)\citenamefont {Ye},
  \citenamefont {Welborn}, \citenamefont {Ricke},\ and\ \citenamefont
  {Van~Voorhis}}]{Ye2018}%
  \BibitemOpen
  \bibfield  {author} {\bibinfo {author} {\bibfnamefont {H.-Z.}\ \bibnamefont
  {Ye}}, \bibinfo {author} {\bibfnamefont {M.}~\bibnamefont {Welborn}},
  \bibinfo {author} {\bibfnamefont {N.~D.}\ \bibnamefont {Ricke}}, \ and\
  \bibinfo {author} {\bibfnamefont {T.}~\bibnamefont {Van~Voorhis}},\ }\href
  {\doibase 10.1063/1.5053992} {\bibfield  {journal} {\bibinfo  {journal} {The
  Journal of Chemical Physics}\ }\textbf {\bibinfo {volume} {149}},\ \bibinfo
  {pages} {194108} (\bibinfo {year} {2018})}\BibitemShut {NoStop}%
\bibitem [{\citenamefont {Pham}\ \emph {et~al.}(2018)\citenamefont {Pham},
  \citenamefont {Bernales},\ and\ \citenamefont {Gagliardi}}]{Pham2018}%
  \BibitemOpen
  \bibfield  {author} {\bibinfo {author} {\bibfnamefont {H.~Q.}\ \bibnamefont
  {Pham}}, \bibinfo {author} {\bibfnamefont {V.}~\bibnamefont {Bernales}}, \
  and\ \bibinfo {author} {\bibfnamefont {L.}~\bibnamefont {Gagliardi}},\ }\href
  {\doibase 10.1021/acs.jctc.7b01248} {\bibfield  {journal} {\bibinfo
  {journal} {Journal of Chemical Theory and Computation}\ }\textbf {\bibinfo
  {volume} {14}},\ \bibinfo {pages} {1960} (\bibinfo {year}
  {2018})}\BibitemShut {NoStop}%
\bibitem [{Note2()}]{Note2}%
  \BibitemOpen
  \bibinfo {note} {We assume that the number of fermions is larger that
  $N_{\mathrm {imp}}$, see Ref.~\protect \rev@citealpnum
  {Klich2006}.}\BibitemShut {Stop}%
\bibitem [{\citenamefont {White}(1992)}]{White1992}%
  \BibitemOpen
  \bibfield  {author} {\bibinfo {author} {\bibfnamefont {S.~R.}\ \bibnamefont
  {White}},\ }\href {\doibase 10.1103/PhysRevLett.69.2863} {\bibfield
  {journal} {\bibinfo  {journal} {Phys. Rev. Lett.}\ }\textbf {\bibinfo
  {volume} {69}},\ \bibinfo {pages} {2863} (\bibinfo {year}
  {1992})}\BibitemShut {NoStop}%
\bibitem [{Note3()}]{Note3}%
  \BibitemOpen
  \bibinfo {note} {Using the relations between the impurity/bath and the
  original fermions, correlations for larger $r$ could in principle be
  evaluated. However, they pick up a contribution from the approximate wave
  function $|\Phi \rangle $ in addition to the one from $|\Phi _{\mathrm
  {imp}}\rangle $. In our case, we have verified that they immediately deviate
  from the correct values, but this is expected since we use the simplest
  possible approximate $\Phi \rangle $. when constructing the impurity
  Hamiltonian.}\BibitemShut {Stop}%
\bibitem [{\citenamefont {Li}\ and\ \citenamefont {Haldane}(2008)}]{Li2008}%
  \BibitemOpen
  \bibfield  {author} {\bibinfo {author} {\bibfnamefont {H.}~\bibnamefont
  {Li}}\ and\ \bibinfo {author} {\bibfnamefont {F.~D.~M.}\ \bibnamefont
  {Haldane}},\ }\href {\doibase 10.1103/PhysRevLett.101.010504} {\bibfield
  {journal} {\bibinfo  {journal} {Phys. Rev. Lett.}\ }\textbf {\bibinfo
  {volume} {101}},\ \bibinfo {pages} {010504} (\bibinfo {year}
  {2008})}\BibitemShut {NoStop}%
\bibitem [{Note4()}]{Note4}%
  \BibitemOpen
  \bibinfo {note} {Comparable results are obtained when $A$ only contains a
  subset of the fragment}\BibitemShut {NoStop}%
\bibitem [{Note5()}]{Note5}%
  \BibitemOpen
  \bibinfo {note} {Notice that in the case $V_2 = 0$, the crossing at $V_1 = 2$
  is dictated by the extra degeneracies coming from the SU(2) symmetry at this
  point, which coincides with the transition point. However, the DET method
  does not conserve this symmetry when constructing the impurity Hamiltonian,
  therefore the crossing we observe is not constrained, and in fact, is
  slightly shifted to larger $V_1\geq 2$.}\BibitemShut {Stop}%
\bibitem [{\citenamefont {Kitazawa}(1997)}]{Kitazawa1997}%
  \BibitemOpen
  \bibfield  {author} {\bibinfo {author} {\bibfnamefont {A.}~\bibnamefont
  {Kitazawa}},\ }\href {http://stacks.iop.org/0305-4470/30/i=9/a=005}
  {\bibfield  {journal} {\bibinfo  {journal} {Journal of Physics A:
  Mathematical and General}\ }\textbf {\bibinfo {volume} {30}},\ \bibinfo
  {pages} {L285} (\bibinfo {year} {1997})}\BibitemShut {NoStop}%
\bibitem [{\citenamefont {Nishimoto}\ and\ \citenamefont
  {Hotta}(2009)}]{Nishimoto2009}%
  \BibitemOpen
  \bibfield  {author} {\bibinfo {author} {\bibfnamefont {S.}~\bibnamefont
  {Nishimoto}}\ and\ \bibinfo {author} {\bibfnamefont {C.}~\bibnamefont
  {Hotta}},\ }\href {\doibase 10.1103/PhysRevB.79.195124} {\bibfield  {journal}
  {\bibinfo  {journal} {Phys. Rev. B}\ }\textbf {\bibinfo {volume} {79}},\
  \bibinfo {pages} {195124} (\bibinfo {year} {2009})}\BibitemShut {NoStop}%
\bibitem [{\citenamefont {Hotta}\ \emph {et~al.}(2006)\citenamefont {Hotta},
  \citenamefont {Furukawa}, \citenamefont {Nakagawa},\ and\ \citenamefont
  {Kubo}}]{Hotta2006a}%
  \BibitemOpen
  \bibfield  {author} {\bibinfo {author} {\bibfnamefont {C.}~\bibnamefont
  {Hotta}}, \bibinfo {author} {\bibfnamefont {N.}~\bibnamefont {Furukawa}},
  \bibinfo {author} {\bibfnamefont {A.}~\bibnamefont {Nakagawa}}, \ and\
  \bibinfo {author} {\bibfnamefont {K.}~\bibnamefont {Kubo}},\ }\href {\doibase
  10.1143/JPSJ.75.123704} {\bibfield  {journal} {\bibinfo  {journal} {Journal
  of the Physical Society of Japan}\ }\textbf {\bibinfo {volume} {75}},\
  \bibinfo {pages} {123704} (\bibinfo {year} {2006})}\BibitemShut {NoStop}%
\bibitem [{\citenamefont {Hotta}\ and\ \citenamefont
  {Furukawa}(2006)}]{Hotta2006b}%
  \BibitemOpen
  \bibfield  {author} {\bibinfo {author} {\bibfnamefont {C.}~\bibnamefont
  {Hotta}}\ and\ \bibinfo {author} {\bibfnamefont {N.}~\bibnamefont
  {Furukawa}},\ }\href {\doibase 10.1103/PhysRevB.74.193107} {\bibfield
  {journal} {\bibinfo  {journal} {Phys. Rev. B}\ }\textbf {\bibinfo {volume}
  {74}},\ \bibinfo {pages} {193107} (\bibinfo {year} {2006})}\BibitemShut
  {NoStop}%
\bibitem [{\citenamefont {Miyazaki}\ \emph {et~al.}(2009)\citenamefont
  {Miyazaki}, \citenamefont {Hotta}, \citenamefont {Miyahara}, \citenamefont
  {Matsuda},\ and\ \citenamefont {Furukawa}}]{Miyazaki2009}%
  \BibitemOpen
  \bibfield  {author} {\bibinfo {author} {\bibfnamefont {M.}~\bibnamefont
  {Miyazaki}}, \bibinfo {author} {\bibfnamefont {C.}~\bibnamefont {Hotta}},
  \bibinfo {author} {\bibfnamefont {S.}~\bibnamefont {Miyahara}}, \bibinfo
  {author} {\bibfnamefont {K.}~\bibnamefont {Matsuda}}, \ and\ \bibinfo
  {author} {\bibfnamefont {N.}~\bibnamefont {Furukawa}},\ }\href {\doibase
  10.1143/JPSJ.78.014707} {\bibfield  {journal} {\bibinfo  {journal} {Journal
  of the Physical Society of Japan}\ }\textbf {\bibinfo {volume} {78}},\
  \bibinfo {pages} {014707} (\bibinfo {year} {2009})}\BibitemShut {NoStop}%
\bibitem [{Note6()}]{Note6}%
  \BibitemOpen
  \bibinfo {note} {Since DMRG favours low entangelement states, it may
  overestimate the stability of the CDW phase}\BibitemShut {NoStop}%
\end{thebibliography}%

\end{document}